\documentclass[twocolumn]{aastex701}
\usepackage{cjhebrew}

\begin{document}

\title{Other red dots: A possible GLIMPSE of normal AGB stars at Cosmic Noon through extreme lensing}

\author[orcid=0000-0001-6278-032X]{Lukas J. Furtak}
\affiliation{Department of Astronomy, The University of Texas at Austin, Austin, TX 78712, USA}
\affiliation{Cosmic Frontier Center, The University of Texas at Austin, Austin, TX 78712, USA}
\affiliation{Department of Physics, Ben-Gurion University of the Negev, P.O. Box 653, Be'er-Sheva 8410501, Israel}
\email[show]{\url{furtak@utexas.edu}}

\author[orcid=0000-0002-0350-4488]{Adi Zitrin} 
\affiliation{Department of Physics, Ben-Gurion University of the Negev, P.O. Box 653, Be'er-Sheva 8410501, Israel}
\email{zitrin@bgu.ac.il}

\author[orcid=0000-0003-1096-2636]{Erik Zackrisson}
\affiliation{Observational Astrophysics, Department of Physics and Astronomy, Uppsala University, Box 516, SE-751 20 Uppsala, Sweden}
\email{erik.zackrisson@physics.uu.se}

\author[orcid=0000-0002-5588-9156]{Vasily Kokorev}
\affiliation{Department of Astronomy, The University of Texas at Austin, Austin, TX 78712, USA}
\affiliation{Cosmic Frontier Center, The University of Texas at Austin, Austin, TX 78712, USA}
\email{vasily.kokorev.astro@gmail.com}

\author[orcid=0000-0003-1282-7454]{Anthony J. Taylor}
\affiliation{Department of Astronomy, The University of Texas at Austin, Austin, TX 78712, USA}
\affiliation{Cosmic Frontier Center, The University of Texas at Austin, Austin, TX 78712, USA}
\email{anthony.taylor@austin.utexas.edu}

\author[0000-0003-2718-8640]{Joseph F. V. Allingham}
\affiliation{Department of Physics, Ben-Gurion University of the Negev, P.O. Box 653, Be'er-Sheva 8410501, Israel}
\email{allingha@post.bgu.ac.il}

\author[orcid=0000-0002-0302-2577]{John Chisholm}
\affiliation{Department of Astronomy, The University of Texas at Austin, Austin, TX 78712, USA}
\affiliation{Cosmic Frontier Center, The University of Texas at Austin, Austin, TX 78712, USA}
\email{Chisholm@austin.utexas.edu}

\author[0000-0001-9065-3926]{Jose M. Diego}
\affiliation{Instituto de F\'isica de Cantabria (CSIC-UC). Avda. Los Castros s/n. 39005 Santander, Spain}
\email{jdiego@ifca.unican.es}

\author[orcid=0000-0002-7570-0824]{Hakim Atek}
\affiliation{Institut d'Astrophysique de Paris, CNRS, Sorbonne Universit\'e, 98bis Boulevard Arago, 75014, Paris, France}
\email{atek@iap.fr}

\author[0000-0001-5538-2614]{Kristen B. W. McQuinn}
\affiliation{Space Telescope Science Institute, 3700 San Martin Drive, Baltimore, MD, 21218, USA}
\affiliation{Rutgers University, Department of Physics and Astronomy, 136 Frelinghuysen Road, Piscataway, NJ 08854, USA}
\email{kmcquinn@stsci.edu}

\author[0000-0003-4564-2771]{Ryan Endsley}
\affiliation{Department of Astronomy, The University of Texas at Austin, Austin, TX 78712, USA}
\affiliation{Cosmic Frontier Center, The University of Texas at Austin, Austin, TX 78712, USA}
\email{ryan.endsley@austin.utexas.edu}

\author[0000-0002-9651-5716]{Richard Pan}
\affiliation{Department of Physics \& Astronomy, Tufts University, MA 02155, USA}
\email{Richard.Pan@tufts.edu}

\author[0000-0003-2680-005X]{Gabriel Brammer}
\affiliation{Cosmic Dawn Center (DAWN), Copenhagen, Denmark}
\affiliation{Niels Bohr Institute, University of Copenhagen, Jagtvej 128, K{\o}benhavn N, DK-2200, Denmark}
\email{gabriel.brammer@nbi.ku.dk}

\author[0000-0001-7232-5355]{Qinyue Fei}
\affiliation{David A. Dunlap Department of Astronomy and Astrophysics, University of Toronto, 50 St. George Street, Toronto, Ontario, M5S 3H4, Canada}
\email{qyfei.astro@gmail.com}

\author[0000-0001-7201-5066]{Seiji Fujimoto}
\affiliation{David A. Dunlap Department of Astronomy and Astrophysics, University of Toronto, 50 St. George Street, Toronto, Ontario, M5S 3H4, Canada}
\affiliation{Dunlap Institute for Astronomy and Astrophysics, 50 St. George Street, Toronto, Ontario, M5S 3H4, Canada}
\email{seiji.fujimoto@utoronto.ca}

\author[0000-0003-4512-8705]{Tiger Y.-Y. Hsiao}
\affiliation{Department of Astronomy, The University of Texas at Austin, Austin, TX 78712, USA}
\affiliation{Cosmic Frontier Center, The University of Texas at Austin, Austin, TX 78712, USA}
\email{tiger.hsiao@utexas.edu}

\author[0000-0003-3142-997X]{Patrick L. Kelly} 
\affiliation{Minnesota Institute for Astrophysics, University of Minnesota, 116 Church St. SE, Minneapolis, MN 55455, USA}
\email{plkelly@umn.edu}

\author[0000-0002-3897-6856]{Damien Korber}
\affiliation{Department of Astronomy, University of Geneva, Chemin Pegasi 51, 1290 Versoix, Switzerland}
\email{damien.korber@protonmail.ch}

\author[0000-0002-7876-4321]{Ashish K. Meena}
\affiliation{Department of Physics, Indian Institute of Science, Bengaluru 560012, India}
\email{ashishmeena766@gmail.com}

\author[0000-0003-3997-5705]{Rohan P.~Naidu}
\affiliation{MIT Kavli Institute for Astrophysics and Space Research, 70 Vassar Street, Cambridge, MA 02139, USA}
\email{rnaidu@mit.edu}

\author[0000-0001-8419-3062]{Alberto Saldana-Lopez}
\affiliation{Department of Astronomy, Oskar Klein Centre, Stockholm University, 106 91 Stockholm, Sweden}
\email{alberto.saldana-lopez@astro.su.se}


\begin{abstract}
We report the discovery of four extremely faint ($m_{\mathrm{F444W}}\gtrsim29$) red point sources in recent ultra-deep JWST/NIRCam images of the strong lensing galaxy cluster Abell~S1063. All four sources sit in lensed arcs, on the symmetry points very close to the critical curves for their host-galaxies' redshifts ($z\sim1-4$). Remarkably, these point sources appear in most arcs that are sufficiently faint close to the critical curve's position ($<21\,\mathrm{nJy}\,\mathrm{arcsec}^{-2}$ in F115W). This suggests that -- unlike previous caustic-crossing events or lensed stars -- thanks to the unprecedented depth of the GLIMPSE observations paired with the extreme lensing magnification (up to $\mu\sim10^4$) we might be resolving the lower-mass ($M\sim1-11\,\mathrm{M}_{\odot}$) red stellar population. Concretely, we detect three likely extremely magnified asymptotic giant branch (AGB) stars ($T_{\mathrm{eff}}\sim3200-3750$\,K), and one yellow super-giant star ($T_{\mathrm{eff}}\sim6750$\,K) -- possibly a yellow hyper-giant or a Cepheid. In addition to offering the first glimpse at low-mass extremely magnified stars, these detections open a possible window into stellar populations, evolution, and chemical enrichment at high redshifts, and could pave the way for using lensed stars such as these as standard candles to populate the distance ladder at cosmological redshifts.
\end{abstract}



\section{Introduction} \label{sec:intro}
The initial stellar mass function \citep[IMF;][]{salpeter55} of galaxies represents one of the best tools to understand stellar physics, populations, and the evolution of galaxies throughout the history of the Universe \citep[e.g.][]{scalo86,bastian10}. While stellar populations and physics are fairly well constrained in our Galaxy and the local Universe \citep[e.g.][]{chabrier03,bastian10,kroupa13}, they remain more uncertain at higher redshifts and recent observations out to $z=0.7$ suggest that the low-mass stars in particular might have been more abundant in the early Universe \citep[e.g.][]{cheng26}. The low-mass stellar population, which makes out the bulk of the stellar mass in galaxies, is difficult to observe directly, even in the vicinity of the Milky Way. Resolving individual stars beyond the local Universe, at cosmological distances, is however normally not possible -- unless assisted by gravitational lensing, specifically the extreme magnifications that occur close to the caustics of a gravitational lens \citep[][]{miralda-escude91}.

Such so-called caustic-crossing events occur when a star in a gravitationally lensed arc crosses the caustics of a typical cluster lens, the critical lines projected into the source plane, where the magnification reaches extreme -- theoretically near infinite -- values\footnote{This is at least the case in the geometric optics approximation, excluding micro-lenses in the cluster.}. The phenomenon is further accentuated by micro-lensing effects from point-masses in the lens \citep[e.g.][]{venmuadhav17,diego18,palencia24}, increasing the rate of events and lowering the attainable magnification to finite values. This effect was predicted by \citet{miralda-escude91} and observed for the first time by \citet{kelly18}. Since then, thanks to the sensitivity and spatial resolution of the \textit{Hubble Space Telescope} (HST) and the JWST \citep[][]{gardner23}, many caustic-crossing or extremely lensed stars have been observed \citep[e.g.][]{chen19,kelly22,meena23a,meena23b,diego23a}, out to high redshifts ($z\sim6$), culminating in the detection of more than 40 caustic-crossing objects at once in one arc \citep{fudamoto25}, and even spectroscopic observations of single stars at cosmological distances \citep[e.g.][]{vanzella20,diego22,furtak24a}. Most caustic-crossing events are transients due to the transverse motion between lens, source, and observer \citep[e.g.][]{kelly18,chen19,fudamoto25} but persistent sources also exist \citep[e.g.][]{meena23b,diego23a}, such as e.g.\ `\textit{Earendel}' at $z=6.2$ \citep{welch22a,welch22b}.

Caustic-crossing events represent the only way to resolve single stars at cosmological distances. While they can be used to constrain dark matter (DM) models \citep[e.g.][]{oguri18,diego18,diego24a,diego24c}, and the intra-cluster light (ICL) of the intervening SL clusters \citep[e.g.][]{diego23b}, they can also be used to populate the Hertzsprung-Russel diagram, to constrain the IMF at high redshifts, and might represent a prime method to directly observe Population~III stars \citep[][]{windhorst18}. A first such attempt to constrain the high-redshift IMF with extremely magnified stars was made by \citet{meena25} who found that the high-mass end of the IMF departs from both the \citet{salpeter55} and \citet{chabrier03} shapes at $z\sim1-3$ (see also \citealt{li25}). Despite the extreme magnifications \citep[typically $\sim10^3-10^4$; see e.g.][]{palencia24}, observed caustic-crossing objects so-far are all rare luminous super-giants, both blue \citep[e.g.][]{kelly18,welch22a,diego22,furtak24a} and red \citep[e.g.][]{diego23a,fudamoto25} and thus only probe the extreme high-mass end of the IMF (i.e.\ $M\gtrsim10\,\mathrm{M}_{\odot}$; e.g.\ \citealt{levesque17}). Low-mass stars on the other hand, remain elusive and have not been observed in extreme lensing events to date (though cf.\ e.g.\ \citealt{leier16}).

In this work, we now report the discovery of several red point-sources located near the symmetry points of arcs lensed by the strong lensing (SL) galaxy cluster Abell~S1063 \citep[AS1063; $z_{\mathrm{d}}=0.348$; e.g.][]{abell89,balestra13,monna14}. These objects differ however from typical lensed stars in several ways: (i) their abundance, being detected in several arcs in the same cluster at once, (ii) their very red colors and smooth black-body-like rest-frame optical slopes, and (iii) that they are most likely persistent sources. As we will develop further throughout this work, we tentatively conclude that most of these lensed stars are most likely asymptotic giant branch (AGB) stars, and thus represent the more `normal', lower-mass ($M=1-11\,\mathrm{M}_{\odot}$), stellar population of their lensed host galaxies at $z\sim1-4$ as opposed to the super-giants usually seen as caustic-crossing objects, which originate from rare high-mass stars. These became observable for the first time here in AS1063 thanks to the unprecedented depth ($\sim31$\,magnitudes) achieved in the near-infrared (NIR) by the recent JWST imaging program GLIMPSE \citep[GO-3293; PIs: H.~Atek \& J.~Chisholm;][]{atek25}.

This work is structured as follows: In section~\ref{sec:data}, we briefly introduce the observations and data used. The results of our analysis are presented in section~\ref{sec:result}, and then discussed in section~\ref{sec:discussion}. Finally, we give a short summary and discuss future prospects in the conclusion in section~\ref{sec:conclusion}. Throughout this work, we assume a standard flat $\Lambda$CDM cosmology with $H_0=70\,\frac{\mathrm{km}}{\mathrm{s}\,\mathrm{Mpc}}$, $\Omega_{\Lambda}=0.7$, and $\Omega_\mathrm{m}=0.3$. Magnitudes are quoted in the AB system \citep{oke83} and uncertainties represent $1\sigma$ ranges unless stated otherwise.

\section{Observations} \label{sec:data}
We primarily use the ultra-deep JWST imaging of AS1063 obtained by the GLIMPSE program with the \textit{Near Infrared Camera} \citep[NIRCam;][]{rieke23}. These observations comprise seven broad-band filters (F090W, F115W, F150W, F200W, F277W, F356W, and F444W) and two medium-band filters (F410M and F480M), and achieve depths down to $31$\,magnitudes at $5\sigma$. The data are reduced and drizzled into mosaics following the prescriptions in \citet{endsley24} using a customized version of the standard NIRCam reduction pipeline, in particular using an improved flat-field calibration which increases the depth by up to 0.5\,magnitudes. For the details of the data reduction, and \textit{bright cluster galaxy} (BCG) subtraction, we refer the reader to the GLIMPSE survey paper, \citet{atek25}. Additional, archival, NIRCam data available are shallower observations from GO-1840 (PI: J.~Alvarez-Marquez) in F115W, F150W, F200W, F250M, F300M, and F444W. These were reduced with the same methods as the GLIMPSE mosaics.

As ancillary data, we make use of the ultra-deep HST observations of AS1063 from the \textit{The Cluster Lensing and Supernova Survey with Hubble} \citep[CLASH; Program ID: 12065; PI: M.~Postman;][]{postman12}, \textit{Hubble Frontier Fields} \citep[HFF; Program ID: 14037; PI: J.~M.~Lotz;][]{lotz17} and \textit{Beyond Ultra-deep Frontier Fields And Legacy Observations} \citep[BUFFALO; Program ID: 15117; PI: C.~L.~Steinhardt;][]{steinhardt20}, programs which comprise imaging with the \textit{Advanced Camera for Surveys} (ACS) in the F435W-, F606W-, and F814W-bands, and with the \textit{Wide-Field Camera Three} (WFC3) in the infrared (IR) bands: F105W, F125W, F140W, and F160W. From HST, we also have WFC3-UVIS F200LP and F350LP imaging at our disposal from the \textit{Flashlights} program \citep[Program ID: 15936; PI: P.~L.~Kelly;][]{kelly22}. These two filters are of particular value since their wide band-passes enable depths comparable to GLIMPSE ($30.5-31$\,magnitudes at $5\sigma$).

The GLIMPSE field of AS1063 was also observed with the \textit{Near Infrared Spectrograph} \citep[NIRSpec;][]{jakobsen22,boeker23} in the framework of the JWST DDT-9223 program (PIs: S.~Fujimoto \& R.~P.~Naidu), targeted at the Population~III candidate by \citet{fujimoto25a}. The micro-shutter array \citep[MSA;][]{ferruit22,jakobsen24} G395M/F295LP configuration of these observations included one of the lensed star candidates reported here, the host galaxy of another, and several multiple images detected in AS1063, with about 9\,h of exposure time. These MSA data were reduced and extracted with \texttt{MSAEXP} \citep[\texttt{v0.9.8};][]{msaexp} and we refer the reader to \citet{fujimoto25b} for more details on the observations, data reduction, and spectral extraction.

\section{Red caustic crossing sources} \label{sec:result}

\begin{deluxetable*}{cccccccc}
\tablecaption{Lensed star candidates and some of their properties. \label{tab:properties}}
\tablehead{
\colhead{ID} & \colhead{$\alpha$ [Degree]} & \colhead{$\delta$ [Degree]} & \colhead{$z$} & \colhead{$m_{\mathrm{F444W}}$} & $\mathrm{F150W}-\mathrm{F444W}$ & \colhead{$\mu_{\mathrm{macro}}$} & \colhead{$d_{\mathrm{crit}}$}
}
\startdata
70878   &   $342.1803171$  &   $-44.5375433$   &   $1.79_{-0.05}^{+0.05}$\tablenotemark{a}   &  $28.84\pm0.05$  &  $3.06\pm0.05$  &   $4767$   & $<0.04\arcsec$\\
43258   &   $342.1797083$  &   $-44.5391325$   &   $3.72$\tablenotemark{b}                   &  $29.67\pm0.10$  &  $1.82\pm0.74$  &   $98$     & $0.62\arcsec$\\
24951   &   $342.1742887$  &   $-44.5308425$   &   $1.43$\tablenotemark{b}                   &  $29.47\pm0.06$  &  $2.08\pm0.62$  &   $332$    & $0.16\arcsec$\\
68501   &   $342.1887413$  &   $-44.5219675$   &   $2.69_{-0.07}^{+0.08}$\tablenotemark{a}   &  $29.65\pm0.08$  &  $2.11\pm0.08$  &   $293$    & $0.20\arcsec$\\
\enddata
\tablenotetext{a}{Geometric redshift based on our SL model of AS1063 (see section~\ref{sec:SL}).}
\tablenotetext{b}{Spectroscopic redshift of the host arc.}
\tablecomments{\textit{Column 1}: ID number in the GLIMPSE catalog (see \citealt{atek25}). \textit{Columns 2 and 3}: Right ascension and declination. \textit{Column 4}: Geometric or spectroscopic redshift. \textit{Column 5}: F444W-band magnitude (see also Tab.~\ref{tab:photometry}). \textit{Column 6}: Rest-frame optical and IR color computed from the photometry in Tab.~\ref{tab:photometry}. \textit{Column 7}: Macro-magnification based on our SL model of AS1063 (see appendix~\ref{app:SL}). \textit{Column 8}: Distance to the critical line, based on our GLIMPSE SL model, at the redshift reported in column~4.}
\end{deluxetable*}

We here report the detection of four red caustic-crossing sources in the GLIMPSE imaging of AS1063, detected serendipitously while cataloging multiply-imaged systems, since they lie in the symmetry points of four multiply-imaged galaxies lensed into arcs. Unlike typical lensed stars, these lensed star candidates are all red ($\mathrm{F150W}-\mathrm{F444W}\sim2-3$), and two of them are possibly persistent since we find them to be detected in observations taken one year prior to GLIMPSE (see appendix~\ref{app:photomety-table}), albeit marginally. All four objects lack a detectable counter-image. This means that they could either lie exactly on the critical line and this in reality be two images merged into one point-source, i.e.\ separated by less than a point-spread-function (PSF) width (0.092\arcsec\ in F277W), or one of the images is magnified or de-magnified above or below the detection threshold by a micro-lens in the cluster. With magnitudes $\gtrsim29$ in the brightest band (F444W), these lensed stars are exceptionally faint and only detectable thanks to the depths achieved by GLIMPSE. Three of them present very smooth spectral energy distributions (SEDs), more consistent with a single red, i.e.\ low-temperature, black-body than a sharp Balmer-break. The combination of all of these properties leads us to tentatively conclude that these are possibly single, extremely magnified red stars, representative of the common lower-mass stellar population of their host galaxies. The fourth is most likely a yellow-supergiant, and was independently detected and identified by \citet{diego26b}, who dubbed it `\textit{Hedorah}'.

\begin{figure*}
    \centering
    \includegraphics[width=\textwidth]{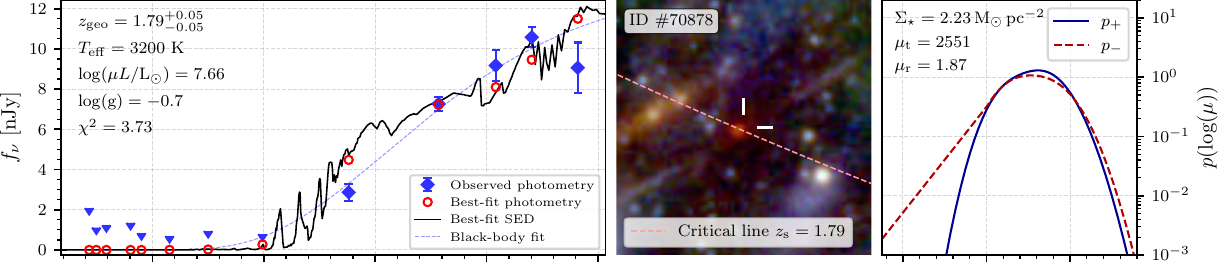}
    \includegraphics[width=\textwidth]{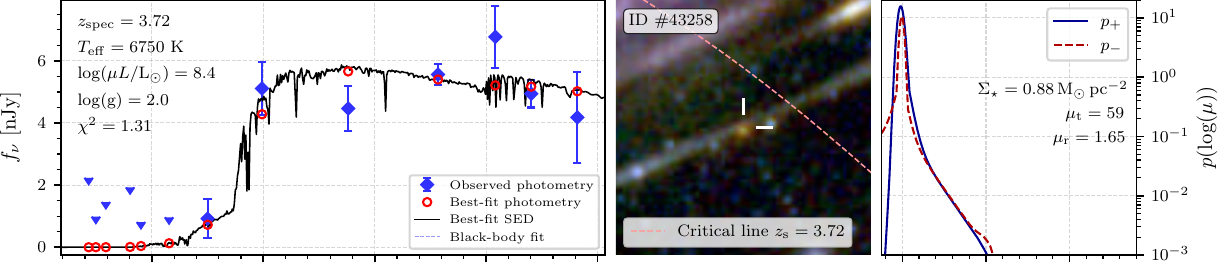}
    \includegraphics[width=\textwidth]{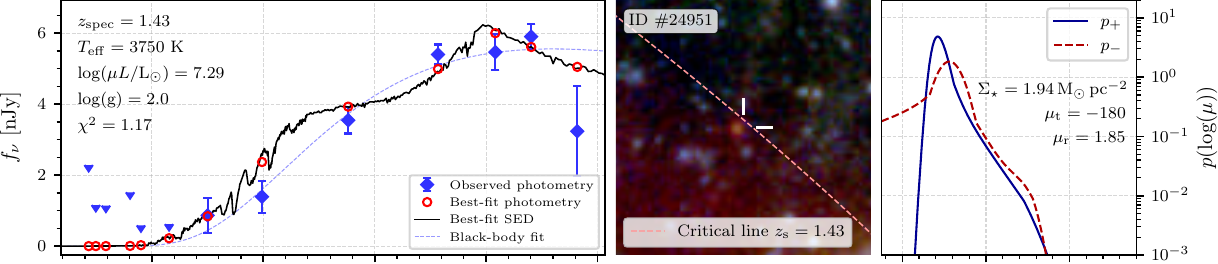}
    \includegraphics[width=\textwidth]{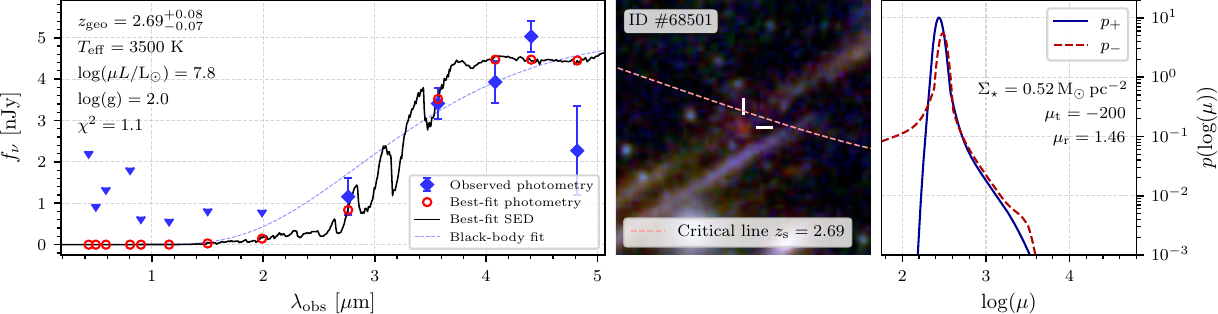}
    \caption{Red lensed star candidates in AS1063. \textit{Left}: Observed deep GLIMPSE JWST and HST photometry (blue; see Tab.~\ref{tab:photometry}), overlaid with the best-fit star SED (black; see section~\ref{sec:SED-modeling}) and the best-fit photometry (red). For the three AGB stars, we overplot the best-fit black-body SED in pale blue. \textit{Middle}: BCG-subtracted $3\arcsec\times3\arcsec$ GLIMPSE JWST composite-color image cutout centered on the lensed star candidates. The critical line from our SL model (see appendix~\ref{app:SL}) for the source redshift of the corresponding object is shown as the pale red line. Each lensed star lies close to the symmetry point of the host arc, very close to the critical line. The host arc of object ID~24951 is too large to fit into the cutout. Note, object ID~43258 is the yellow-supergiant \textit{Hedorah} which was independently identified by \citet{diego26b}. \textit{Right}: Total, combined macro- and micro-lensing magnification probability density functions for each object computed using the \citet{palencia24} analytical approximation, the \citet{beauchesne25b} ICL map to estimate the micro-lens surface mass density $\Sigma_{\star}$, and assuming the cluster-scale macro-magnifications from our SL model listed in Tab.~\ref{tab:properties}. Probability densities for either side of the critical line (positive and negative parity) are shown in blue and red respectively.}
    \label{fig:observations}
\end{figure*}

We measure simple aperture photometry for these lensed star candidates, the details of which are given in appendix~\ref{app:photomety-table}. The host-arc of one of the objects, ID~24951 which is located in multiple-image system~6, has a ground-based spectroscopic redshift measurement ($z_{\mathrm{s}}=1.43$; see e.g.\ \citealt{bergamini19,richard21,beauchesne24}). For the host of ID~43258 (\textit{Hedorah}), labeled multiple-image system~401, we measure a spectroscopic redshift of $z_{\mathrm{s}}=3.7152\pm0.0001$ from the JWST-DDT NIRSpec data (see section~\ref{sec:data}), as detailed further in appendix~\ref{app:spectroscopy}. For the other two (IDs~70878 and~68501), we derive geometric redshifts from our SL model of AS1063 (see section~\ref{sec:SL} and appendix~\ref{app:SL}). The lensed star candidate ID~70878 was also included in the NIRSpec MSA mask, but we did not detect any emission lines or continuum from it. This is however not surprising since the NIRSpec G395M spectrum is not deep enough to detect a $29$\,magnitude continuum, and even though this object is one of the marginal detections in the prior data (GO-1840; see appendix~\ref{app:photomety-table}), we cannot entirely rule-out that it is a transient which might have disappeared by the time the GLIMPSE NIRSpec DDT program was observed. All four objects, along with their SEDs and magnification estimates, are shown in Fig.~\ref{fig:observations} and their observed properties are summarized in Tab.~\ref{tab:properties}. We estimate their gravitational magnifications in section~\ref{sec:SL}, and perform an SED-modeling analysis in section~\ref{sec:SED-modeling}.

\subsection{Gravitational lensing} \label{sec:SL}
We use a new parametric SL model of AS1063, constructed for GLIMPSE with \texttt{AstroLensPy} \citep{allingham26}, the \texttt{python} version of the \citet{zitrin15a} analytic method, to compute the macro-magnifications of the lensed stars and to constrain the geometric redshifts of objects without a spectroscopic redshift measurement. The model is constrained with 91 multiple images of 34 sources, 31 of which have spectroscopic redshifts, and achieves an average image reproduction error of $\Delta_{\mathrm{RMS}}=0.32\arcsec$. We include five new spectroscopic multiple-image redshifts measured from the JWST/NIRSpec DDT program (see section~\ref{sec:data}), as well as, for the first time, constraints on the north-eastern substructure of AS1063 \citep[e.g.][]{bergamini19,richard21}. We refer the reader to appendix~\ref{app:SL} for more details on the SL model and its constraints.

Since the exact position of the critical lines is uncertain by several to a few tens of pixels ($0.04\arcsec/\mathrm{pix}$), the macro-magnification of the lensed stars reported in this work is also highly uncertain. We measure the distance of each object to the critical line, derived from our SL model, at its source redshift reported in Tab.~\ref{tab:properties}, finding $d_{\mathrm{crit}}<0.04\arcsec$ ($<1\,\mathrm{pix}$), $0.62\arcsec$, $0.16\arcsec$, and $0.2\arcsec$ respectively for IDs~70878, 43258, 24951, and~68501. According to the SL model, the macro-magnifications are $\mu_{\mathrm{macro}}\sim4767$, $\sim98$, $\sim332$, and $\sim293$ at the position of each object. The exact positions of the critical lines are however not known, which means that the macro-magnifications are also very uncertain and could be much higher. The magnification gradient across the F277W PSF area\footnote{We use the F277W-PSF for this estimate, because it is the smallest that yields a robust detection of all four objects.}, in the tangential direction, is quite severe, with $\nabla_{\mathrm{t}}\mu\sim1500\,\mathrm{pix}^{-1}$. We note that varying the redshift of object ID~68501 within the $1\sigma$-range of its geometric redshift estimate brings its macro-magnification up to $\mu_{\mathrm{macro}}\sim2600$. Furthermore, the total magnification of our lensed star candidates is also affected by micro-lenses present in the cluster \citep[e.g.][]{diego18,palencia24}. We therefore use an analytic approximation by \citet{palencia24} to compute the magnification probability density as a function of the macro-magnification and the ICL surface mass density at the positions of the lensed stars. The micro-lens surface mass density is computed from an ICL map of AS1063, measured and converted to $\mathrm{M}_{\odot}\,\mathrm{pc}^{-2}$ by \citet{beauchesne25b} using an SED-fitting analysis. The resulting magnification probability densities are shown in the right-hand panels of Fig.~\ref{fig:observations}. Due to the uncertainty in the exact critical lines of the macro SL model, and the stochasticity of micro-lensing events, the actual magnifications are not known and could in theory reach as high as several $10^4$.

\subsection{SED modeling} \label{sec:SED-modeling}

\begin{deluxetable}{cccccc}
\tablecaption{Lensed star candidate SED modeling results. \label{tab:SED_result}}
\tablehead{
\colhead{ID} & \colhead{$z_{\mathrm{fit}}$} & \colhead{$T_{\mathrm{eff}}$ [K]} & \colhead{$\log(\mu L/\mathrm{L}_{\odot})$} & \colhead{$\log(g)$} & \colhead{$\chi^2$}
}
\startdata
70878   &   $1.84$  &    $3200$  &   $7.66$  &   $-0.7$ &   $3.73$\\
43258   &   -       &    $6750$  &   $8.40$  &   $2.00$ &   $1.31$\\
24951   &   -       &    $3750$  &   $7.29$  &   $2.00$ &   $1.17$\\
68501   &   $2.74$  &    $3500$  &   $7.80$  &   $2.00$ &   $1.10$\\
\enddata
\tablecomments{The fitted redshift solutions for IDs~70787 and~68501 lie within the $1\sigma$-range of their geometric redshifts (see Tab.~\ref{tab:properties}). The sixth column contains the reduced $\chi^2$ for each fit.}
\end{deluxetable}

To explore the possibility that the lensed point sources detected are individual, highly magnified stars, we fit the HST and JWST photometric fluxes with the stellar SED templates of \citet{lejeune97}. We adopt the spectroscopic redshifts of the host-arcs in the case of objects ID~43258 and~24951, and allow the redshift to vary within $1\sigma$ around their geometric redshift estimates for the other two objects (see Tab.~\ref{tab:properties}). We moreover neglect the effects of dust along the line of sight, since the contributions of circum-stellar and inter-stellar dust cannot easily be separated in this situation, and limit the fits to surface gravities $\log(g)\leq2.0$, as appropriate for giants, which are the stars that are luminous enough to be detected, and metallicities $[\mathrm{M/H}]=0$. Note that varying the metallicity does not significantly affect the results of the SED-fits, which suggests that the available data (broad-band photometry) are not sensitive to the stellar metallicity. The fits are then based on either three or four free parameters: the effective temperature $T_\mathrm{eff}$, the surface gravity $\log(g)$, the magnification-boosted bolometric luminosity $\mu L_\mathrm{bol}/L_\odot$, and the redshift for the two objects without spectroscopic redshifts. The best-fit parameters and reduced $\chi^2$ of all four objects are given in Tab.~\ref{tab:SED_result} and shown in Fig.~\ref{fig:observations}.

For three of the objects (IDs~24951, 68501, and~70878), the best-fitting (reduced $\chi^2=1.10-3.73$) $T_\mathrm{eff}$ is in the range $\approx3200-3750$\,K, with $\log(\mu L_\mathrm{bol}/\mathrm{L}_{\odot})\approx 7.3-8.0$. Provided that the magnifications are close to $\mu\sim10^{3}$, the intrinsic luminosity would be in the range $\log(L_\mathrm{bol}/\mathrm{L}_{\odot})\sim4.3-5.0$. These parameters are broadly consistent with AGB stars. In the case of substantial dust reddening/extinction, the inferred $T_\mathrm{eff}$ and $L_\mathrm{bol}$ would both increase, thereby shifting the stars into the territory of intrinsically brighter, higher-$T_\mathrm{eff}$ supergiants. As an example, the best-fit $T_\mathrm{eff}=3750$\,K of object ID~24951 would shift to 4000\,K (4500\,K) if subject to 1.0\,magnitudes (2.0\,magnitudes) of rest-frame $V$-band extinction at the redshift of the source, assuming a Milky-Way-like extinction curve.

Object ID~43258 on the other hand is best-fit (reduced $\chi^2=1.31$) with $T_\mathrm{eff}\approx6750$\,K and $\log(\mu L_\mathrm{bol}/\mathrm{L}_{\odot})\approx8.4$. For $\mu\sim10^{3}$, this would place the object at a luminosity of $\log(L_\mathrm{bol}/\mathrm{L}_\odot)\sim5.4$ and indicates that the object might be a yellow super-giant and perhaps even a Cepheid on the bright end of its luminosity cycle, in agreement with the results by \citet{diego26b}.

\section{Discussion} \label{sec:discussion}
As illustrated in section~\ref{sec:SED-modeling}, the caustic-crossing sources presented in this work are best fit with AGB stars, and a yellow super-giant in the case of ID~43258. We will discuss these results and their implications in the following, the AGB star candidates in section~\ref{sec:AGB-stars}, and the yellow super-giant `\textit{Hedorah}' \citep{diego26b} in section~\ref{sec:cepheid}. Finally, we briefly discuss alternative interpretations of these sources in section~\ref{sec:contamination}. 

\subsection{Lensed AGB stars at $z\sim1-3$} \label{sec:AGB-stars}

\begin{figure*}
    \centering
    \includegraphics[width=\textwidth]{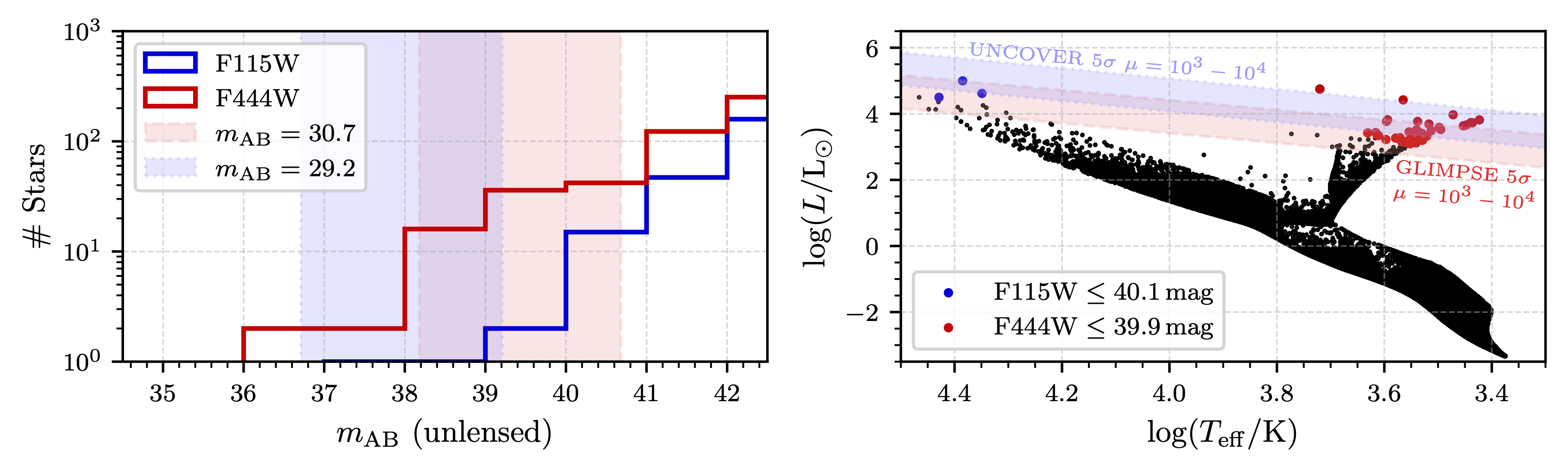}
    \caption{Simulated brightness properties of a 4.5\,Gyr old, constant SFH stellar population of $M_{\star}=6\times10^5\,\mathrm{M}_{\odot}$ at $z=1.43$ (the spectroscopic redshift of ID~24951). This represents the most conservative population in the vicinity of the lensed star since it corresponds to the age of the Universe at that redshift. \textit{Left}: Distribution of apparent magnitudes in F115W and F444W for the brightest stars in the simulated population. The red-shaded area represents magnitudes corresponding to the $5\sigma$ GLIMPSE depth reported in \citet{atek25} and magnifications spanning $\mu=10^3-10^4$. The blue-shaded area shows the same for the next shallower JWST-observed SL field, UNCOVER \citep[][]{weaver24}. \textit{Right}: Hertzsprung-Russel diagram of the simulated stellar population (black) with stars detectable at GLIMPSE depths ($5\sigma$; assuming a magnification of $\mu=5000$) in F444W and F115W, highlighted as red and blue dots respectively. Note that the majority of stars bright enough to be detected lie on the AGB. The red and blue shaded areas again represent the GLIMPSE and UNCOVER detection limits for magnifications spanning $\mu=10^3-10^4$. This illustrates that, at GLIMPSE depth, extreme lensing events become sensitive to AGB star luminosities.}
    \label{fig:HR-diagram}
\end{figure*}

The three AGB stars, IDs~70878, 24951, and~68501, with $T_{\mathrm{eff}}=3200-3750$\,K strongly differ from typical extremely magnified stars since they are colder, and if indeed on the AGB, of much lower mass ($M\sim1-11\,\mathrm{M}_{\odot}$; e.g.\ \citealt{doherty17}), than the more luminous super- and hyper-giants typically detected through extreme lensing in shallower (and bluer) observations \citep[e.g.][]{kelly22,fudamoto25,meena25}. They are therefore representative of the more `normal', intermediate mass, stellar population which is more abundant and long-lived than the massive, short-lived and rare super-giants. Three robust detections in only one observation strongly suggest that the detection of extremely magnified AGB stars becomes a common occurrence provided the depth is sufficient. To illustrate this, we use $Z=0.0152$ stellar isochrones from CMD \texttt{v.3.8}\footnote{Available at: \url{http://stev.oapd.inaf.it/cmd}.} and $[\mathrm{M}/\mathrm{H}]=0$ stellar SEDs from \citet{lejeune97} to simulate an $M_{\star}=6\times10^5\,\mathrm{M}_{\odot}$ stellar population at $z=1.43$ (the spectroscopic redshift of ID~24951) assuming a constant star-formation history (SFH) and an age of 4.5\,Gyr (i.e.\ the age of the Universe at that redshift). This represents the most conservative stellar population in the host galaxy possible, assuming an arbitrary field of stars. As can be seen in Fig.~\ref{fig:HR-diagram}, with typical total magnifications of $\mu\sim10^3-10^4$ \citep[e.g.][]{palencia24}, and the GLIMPSE depth of $\sim31$\,magnitudes, we are sensitive to AGB star luminosities. In fact, under these assumptions, the number of detectable red (AGB) stars in the F444W-band \emph{exceeds} the number of detectable blue stars in the F115W-band. This suggests that, at GLIMPSE depth, AGB stars become the most commonly detectable lensed stars, unless the highly magnified area happens to probe a recent star-formation region in the host arc, as we discuss more quantitatively further below. The next-deepest JWST-imaged SL field after GLIMPSE is Abell~2744 (A2744) observed in the \textit{Ultradeep NIRSpec and NIRCam Observations before the Epoch of Reionization} \citep[UNCOVER;][]{bezanson24} survey which reached $5\sigma$-depths of 29.2\,magnitudes \citep[][]{weaver24}, also shown in Fig.~\ref{fig:HR-diagram}. While the UNCOVER depth is just sufficient to, in theory, detect the very brightest AGB stars at $z\sim1-2$, we did not find any similar sources in those observations. That could however also be due to the fact that Abell~2744 is a particularly ill-suited cluster for this kind of search since it only hosts very few caustic-crossing arcs, due to the lensing configuration, i.e.\ just the chance alignment of lens and background sources, and none with sufficiently low surface-brightness to achieve sufficient signal-to-noise of such faint embedded sources. Other JWST SL observations so-far are shallower and thus sensitive to brighter stars, in the super- and hyper-giant regime.

As mentioned in section~\ref{sec:SED-modeling}, our SED-modeling cannot entirely rule-out a red super-giant fit, even though that would be slightly hotter ($T_{\mathrm{eff}}\sim4000$\,K) than the AGB star solutions. Red super-giants are evolutionary phases of massive stars ($M\gtrsim15\,\mathrm{M}_{\odot}$) which are short-lived objects and thus only exist in currently or recently star-forming regions of a galaxy. In Fig.~\ref{fig:observations}, it can be seen that the host-arcs of the three lensed AGB stars are extremely faint adjacent to the stars. More quantitatively, we measure the surface-brightnesses of the arcs adjacent to the stars and find them to lie below the detection limit (see appendix~\ref{app:photomety-table}), i.e. $<21\,\mathrm{nJy}\,\mathrm{arcsec}^{-2}$ at $5\sigma$ in the F115W-band. This surface brightness upper-limit allows us to place an upper-limit on the maximum luminosity in the arc in the vicinity of the lensed stars. If we attribute this maximum luminosity to a single luminous star, this places an upper-limit on the main-sequence turn-off in the stellar population, as all present blue stars must be less luminous than this upper-limit. Using the GLIMPSE depth, assuming a magnification of $10^3$, and a mass-luminosity relation of $L/\mathrm{L}_{\odot}=1.4\,(M/\mathrm{M}_{\odot})^{3.5}$ \citep[e.g.][]{yakut07,kippenhahn13}, the non-detections of the host-arcs adjacent to the lensed stars suggest main sequence turn-over masses $M\lesssim12-14\,\mathrm{M}_{\odot}$, and $M\lesssim14-16\,\mathrm{M}_{\odot}$ if the magnification is as low as a few hundred. Stellar evolution tracks suggest that it takes $\gtrsim20$\,Myr for these masses to leave the main sequence, in which case many super-giants would have already disappeared. The majority of stars remaining that are luminous enough to be detected would be AGB stars. This does not necessarily exclude \emph{all} red super-giants across the board however, since they can have progenitor masses down to $M\sim8-10\,\mathrm{M}_{\odot}$ \citep[e.g.][]{levesque17}, and even as low as $M=6\,\mathrm{M}_{\odot}$ according to more recent observations \citep{yang24}. There is a significant overlap between the low-mass end of red super-giants and the most massive AGB and super-AGB stars which can be as massive as $M\sim11\,\mathrm{M}_{\odot}$ \citep[e.g.][]{doherty14,doherty17,gil-pons18}. While the AGB phase of any \emph{individual} star is as short-lived as a red super-giant \citep[$\sim10^4-10^7$\,yr; e.g.][]{vassiliadis93,marigo08,doherty14}, on the level of a population, AGB stars remain for much longer and are thus more likely to be detected in extreme lensing events, provided the observations are deep enough. This in particular is the case since the lower-mass progenitors of AGB stars, are more numerous than the more massive progenitors of red super-giants to begin with.

We therefore here tentatively conclude that these three faint, red, extremely magnified sources detected in AS1063 are most likely extremely magnified AGB stars. Ultimately, the available observations however cannot entirely rule-out red super-giants and other alternative explanations (see section~\ref{sec:contamination} for further discussion). Given the faintness of these sources, it would indeed be very difficult to confirm their true nature with spectroscopy. This is especially true since AGB stars typically do not show emission lines which means that we would need to detect their absorption features instead to measure e.g.\ $\log(g)$ and $T_{\mathrm{eff}}$. As shown and discussed in \citet{lundqvist24}, a detection of stellar absorption lines in extremely lensed stars beyond 28\,magnitudes in a reasonable amount of time is unlikely with JWST and the upcoming \textit{Extremely Large Telescope} (ELT), in particular for stars as cold as $T\sim3500$\,K (see Tab.~\ref{tab:properties}). Instead, future ultra-deep NIRCam imaging programs of SL galaxy clusters that achieve GLIMPSE depths (31\,magnitudes and beyond) will enable us to verify if these faint and red extremely magnified stars are a commonly detectable phenomenon. With larger samples, it will be possible to study the population-level properties of their host arcs and link them to the AGB star detections, as well as constrain the intermediate-mass IMF done in \citet{meena25} for the high-mass IMF. This would enable us to verify if indeed the lower-mass end of the IMF becomes more prominent at high redshifts than in the local Universe as suggested by the findings of \citet{cheng26}.

This possible detection of AGB stars at $z\sim1-3$, at Cosmic Noon, has the potential to open a new window into the study of stellar evolution and chemical enrichment in the Universe. Indeed, metal-poor (i.e.\ Population~\textsc{ii}) AGB stars represent the main contributors to the creation of nitrogen and dust in the Universe \citep[e.g.][]{karakas14,kobayashi20,ventura21}. These lensed AGB star candidates at cosmological distance are therefore particularly interesting in the light of recent JWST discoveries of nitrogen over-abundances in early galaxies from $z\sim2$ out to the highest redshifts \citep[e.g.][]{bunker23,topping24,topping25,arellano-cordova25,cataldi25,berg26}. Carbon-rich AGB stars can also be used as standard candles to directly measure cosmology-independent distances \citep[e.g.][]{freedman20,ripoche20,zgirski21}, which would make them particularly interesting as lensed stars at cosmological distances to expand the distance ladder \citep[see][]{diego26a}, as will be discussed further in section~\ref{sec:cepheid}.

\subsection{A lensed yellow super-giant at $z_{\mathrm{spec}}=3.72$} \label{sec:cepheid}

\begin{figure*}
    \includegraphics[width=\textwidth]{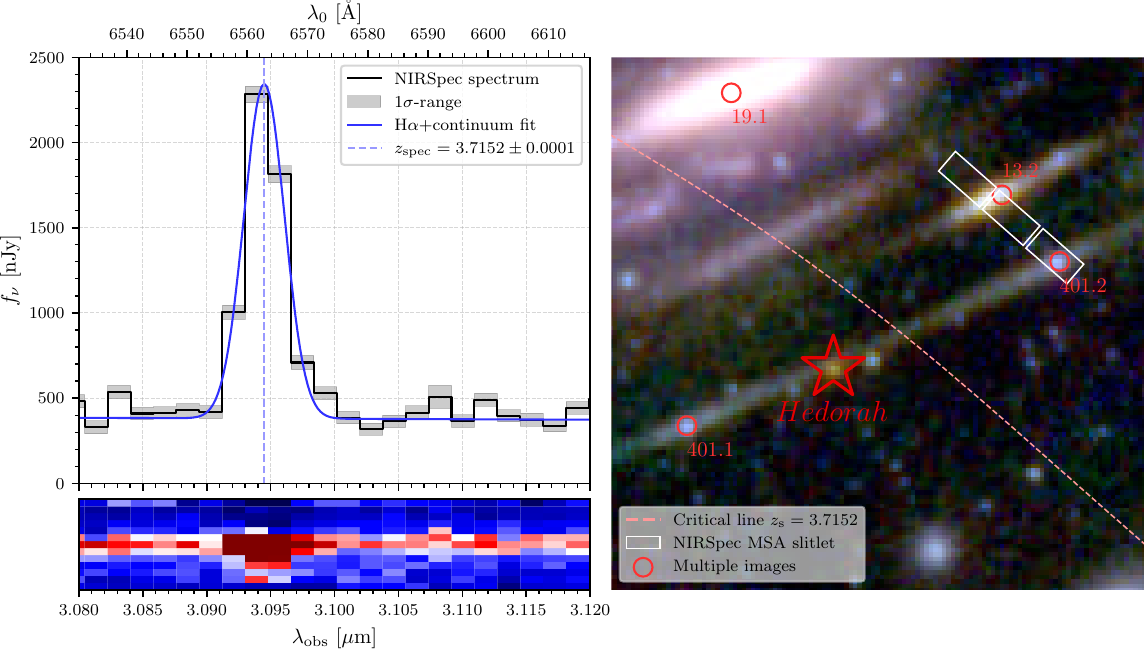}
    \caption{JWST/NIRSpec G395M/F290LP spectroscopy of multiple image system~401, the host arc of ID~43258 \textit{Hedorah}. \textit{Left}: Zoom-in on the H$\alpha$ line, measured at $z_{\mathrm{spec}}=3.7152\pm0.0001$. The identification of this single emission line as H$\alpha$ is facilitated by a multitude of close-by spectroscopically confirmed multiple image systems which places a strong geometric prior on the redshift of system 401. \textit{Right}: BCG-subtracted $5\arcsec\times5\arcsec$ GLIMPSE JWST composite-color image cutout centered on \textit{Hedorah} and its host arc. The critical line from our SL model for the redshift of the arc is shown as the pale red line, known multiple images in AS1063 are marked as red circles, and the NIRSpec MSA slitlets used here are shown in white. The other object in the slit, multiple image number 13.2, is a known Lyman-alpha emitter at $z=4.11$ detected with VLT/MUSE \citep{bergamini19,richard21}.}
    \label{fig:hedorah-host_spec}
\end{figure*}

The yellow super-giant, ID~43258 (\textit{Hedorah}) with $T_{\mathrm{eff}}=6750$\,K, while not an AGB star, is also a unique object as only very few yellow super-giants have been observed as caustic-crossing objects to date \citep[][]{diego23c,furtak24a,nabizadeh25}. This particular object was independently detected and identified by \citet{diego26b}, who also used the GLIMPSE data and also interpret it as most likely a yellow super-giant, though be it at a slightly lower redshift ($z_{\mathrm{geo}}=3.086$) and temperature ($T_{\mathrm{eff}}=6500$\,K). We here measure the spectroscopic redshift of its host arc to be $z_{\mathrm{s}}=3.7152\pm0.0001$, using the H$\alpha$ emission line detected in the JWST/NIRSpec DDT observations shown in Fig.~\ref{fig:hedorah-host_spec}, and include it as a constraint in our SL model (see appendix~\ref{app:SL}). The details of the spectrum extraction and analysis can be found in appendix~\ref{app:spectroscopy}. At $0.62\arcsec$ from the critical line in our SL model, \textit{Hedorah} is among the furthest caustic-crossing objects from the critical line detected to date \citep[e.g.][]{meena23b}. In \citet{diego26b}, it is located even further away ($0.9\arcsec$) due to a different lens model and the lower redshift. With a relatively low macro-magnification ($\mu_{\mathrm{macro}}\sim100$), it requires significant micro- or milli-lensing to explain the absence of a counter-image and boost the emission of a single star at $z=3.72$ to detectable magnitudes: As discussed in great detail in \citet{diego26b}, \textit{Hedorah} is most likely a luminous yellow hyper-giant (a massive $M\sim20-60\,\mathrm{M}_{\star}$ evolving between blue and red hyper-giant phases), or possibly a fainter, variable Cepheid if the magnification is higher ($\mu>2000$). Being slightly closer to the critical line in our model makes the Cepheid scenario slightly more likely, because it increases the probability for a higher total magnification, but overall our results agree with \citet{diego26b}. The blue rest-frame ultra-violet (UV) continuum and H$\alpha$ emission in \textit{Hedorah}'s host arc suggest a young, heavily star-forming environment which could host massive yellow hyper-giants.

Both the yellow hyper-giant and the Cepheid interpretations make \textit{Hedorah} a uniquely interesting object. Yellow hyper-giants are rare transitioning phases of massive stars that will finish their lives as Wolf-Rayet stars and type~II supernovae (SNe). They typically emit relatively strong emission lines \citep[e.g. H$\alpha$, H$\beta$, and more; e.g.][]{jones93,humphreys02}, which could be detectable with JWST or ELT in extremely magnified stars like \textit{Hedorah}, provided the magnification is high enough. Such observations would enable us to directly observe and study SN progenitors beyond the local Universe \citep[e.g.][]{vandyk26}. The fainter Cepheids on the other hand are also of particular interest, especially at cosmological distances, because of their well-known pulsation period-luminosity relation \citep[e.g.][]{leavitt12}, which enables us to measure model-independent distances if the pulsation light-curve can be measured \citep[see e.g.][for a review]{freedman24}. As discussed in \citet{diego26a}, lensed Cepheids, carbon-rich AGB stars, and tip of the red-giant branch (TRGB) stars \citep[e.g.][]{mould86,lee93,mcquinn19} are cosmological standard candles and thus have the potential to expand the cosmic distance ladder by orders of magnitude, even beyond the current reach of type~Ia supernovae. Similarly to the AGB stars discussed in section~\ref{sec:AGB-stars}, we in general become sensitive to Cepheid luminosities \citep[up to $L\sim10^5\,\mathrm{L}_{\odot}$; e.g.][]{turner10} in caustic-crossing events at GLIMPSE depth (see Fig.~\ref{fig:HR-diagram}). This means that future ultra-deep (31\,magnitudes and beyond) JWST imaging observations of SL clusters could detect more Cepheids, AGB, and TRGB stars at different redshifts, allowing us to populate the distance ladder and measure cosmological distances independently of cosmological models \citep{diego26a}.

\subsection{Alternative interpretations} \label{sec:contamination}

\begin{figure}
    \centering
    \includegraphics[width=0.5\textwidth]{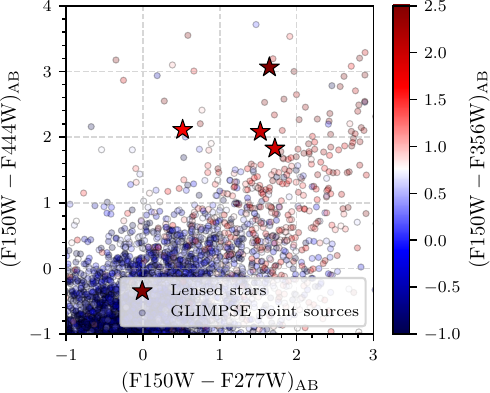}
    \caption{Color-comparison of the lensed star candidates (stars) with point-sources in the GLIMPSE catalog (circles). The lensed stars are less red in $\mathrm{F150W}-\mathrm{F277W}$ than most point-sources of similar $\mathrm{F150W}-\mathrm{F444W}$ color in the catalog due to their smoothly rising red rest-frame optical slopes (see Fig.~\ref{fig:observations}) as opposed to the sharper Balmer-breaks observed typically in galaxies and star-clusters.}
    \label{fig:color}
\end{figure}

Despite having only broad-band photometry and two medium bands at our disposal, the sources reported here have several unique features that strongly disfavor most alternative interpretations to lensed stars. While they do satisfy typical `Little Red Dot' \citep[LRD;][]{matthee24} selection criteria \citep[e.g.][]{kokorev24a}, including the brown-dwarf exclusion color-cut by \citet{greene24}, any higher-redshift objects, such as LRDs, high-redshift galaxies or star-clusters, at the same locations would necessarily be multiply-imaged by the cluster since they would lie \emph{within} the caustics for any source-redshift above the values reported in Tab.~\ref{tab:properties}. This leaves the possibility of chance alignments with faint and red foreground objects such as e.g.\ brown dwarves in the Milky Way (though unlikely since they satisfy the brown-dwarf exclusion color-cuts mentioned above), or compact red star-clusters along the line-of-sight at lower redshift. To assess the probability of a chance alignment, we use the GLIMPSE catalog \citep[][also see section~\ref{sec:data}]{atek25} to search for point-sources of similar color to our lensed star candidates. As can be seen in Fig.~\ref{fig:color}, the four lensed star candidates occupy a distinct region in the color-color space ($\mathrm{F150W}-\mathrm{F444W}$ vs.\ $\mathrm{F150W}-\mathrm{F277W}$), apart from the bulk of the point-source population. Taking the entire population of point-sources in the field, we find a density of $0.7\,\mathrm{arcsec}^{-2}$, which gives a 0.5\,\% probability of finding a random point-source within one F277W PSF width from a critical line. Given the dearth of similarly-colored point-sources across the field, a chance alignment of foreground objects with the critical lines of AS1063 is therefore unlikely.

Finally, there is the possibility of other red objects within the host-galaxy: e.g.\ red super- or hyper-giants, or globular clusters. The possibility of red super-giants was already discussed in detail in section~\ref{sec:AGB-stars}, where we concluded that we cannot definitely rule it out. On the lower magnification end of the distributions shown in Fig.~\ref{fig:observations} ($\mu\lesssim10^2$), these objects could alternatively be compact globular clusters residing in the outskirts of the host galaxies. This scenario is a commonly discussed alternative solution for extremely magnified stars, in particular for persistent ones \citep[e.g.][]{furtak24a,diego23c,diego26b}. In the case of the three lensed AGB star candidates discussed here, the complete absence of any rest-frame UV emission as well as the smooth red slope would require a significant amount of dust attenuation. This is however atypical for mature globular clusters \citep[e.g.][]{bastian14} and therefore disfavors a globular cluster solution. Even if they had the age of the Universe at $z\sim1-3$, globular cluster stellar populations would still be governed by G-type stars and thus show significant rest-frame UV and blue optical emission. In addition, the absence of a counter-image (see Fig.~\ref{fig:observations}) is much harder to explain for a globular cluster, requiring de-magnification by a rare and fortuitously placed milli-lens \citep[e.g.][]{diego23c,diego26b}, than for a single lensed star.

\section{Conclusion} \label{sec:conclusion}
We report the detection of four faint, extremely red point-sources located at the symmetry points of four caustic-crossing arcs in the SL galaxy cluster AS1063. Lying at redshifts $z\sim1-4$, with typical colors of $\mathrm{F150W}-\mathrm{F444W}\sim2-3$, these extremely faint objects ($\gtrsim29$\,magnitudes in the brightest band) were identified in the deepest JWST/NIRCam imaging taken to date, the cycle~2 GLIMPSE program (Program-ID: GO-3293; PIs: H.~Atek \& J.~Chisholm), which achieved observational depths of 31\,magnitudes at $5\sigma$. All four objects are detected as single red point-sources on top the symmetry points of four lensed arcs, very close to and in one case even on top of the critical lines ($d_{\mathrm{crit}}\sim0.04\arcsec-0.6\arcsec$) where the magnifications can reach extreme values (up to $\mu\sim10^4$). Two of our objects are marginally detected in previous, shallower data, but at this stage, we cannot definitely determine whether these are transient or persistent sources. The very red SEDs and smooth rest-frame optical slopes of three of these lensed stars are consistent with single black-body emission rather than a Balmer-break. A more detailed SED-analysis reveals three of them to be best-fit with AGB star templates ($T_{\mathrm{eff}}\sim3200-3750$\,K). The fourth is consistent with a yellow super-giant ($T_{\mathrm{eff}}=6750$\,K), in agreement with the results by \citet{diego26b}, who dubbed this object `\textit{Hedorah}'. We for the first time measure the spectroscopic redshift of \textit{Hedorah}'s host galaxy with JWST/NIRSpec, at $z_{\mathrm{spec}}=3.7152\pm0.0001$. While we find a chance-alignment of the host arcs' critical lines with higher or lower-redshift interlopers to be unlikely, it is however not possible to definitively rule-out other interpretations with the data available at the present time.

We here conclude that the four red extremely magnified objects observed in AS1063 are indeed single lensed stars, three AGB stars and one yellow super-giant, possibly a Cepheid. The red and yellow stars reported here are particularly interesting because they represent a much lower mass, and thus more common, stellar population than the super- and hyper-giants typically seen as caustic-crossing events, and could potentially pave the way for model-independent distance measurements with lensed stars in the future, as suggested by \citet{diego26a}. The relatively high occurrence rate, four detections in one SL cluster, suggest that such red extreme lensing events are a common phenomenon provided observations are deep enough. While spectroscopic confirmation and follow-up observations of these lensed AGB stars and yellow super-giant are unlikely in the near future due to their extreme faint magnitudes and red wavelengths, future ultra-deep JWST/NIRCam imaging of SL galaxy clusters will reveal more such sources once sufficient depth is achieved. Future deep follow-up NIRCam imaging of AS1063 would in particular be required to confirm if some of these extremely lensed stars, most prominently ID~70878, are persistent sources and not transients. This first direct and resolved detection of AGB stars at $z\sim1-3$, and a possible Cepheid at $z=3.72$, opens a new window into stellar physics, abundances and chemical enrichment at high redshifts and represents a further step in populating the Hertzsprung-Russel diagram, and the cosmic distance ladder, at cosmological distances.

%% Please use the acknowledgment and contribution environments. This will 
%% be anonomyized when the "anonymous" style option is used. 
\begin{acknowledgments}
L.J.F. and V.K. acknowledge support from the University of Texas at Austin Cosmic Frontier Center. Support for JWST program GO-3293 was provided by NASA through a grant from the Space Telescope Science Institute (STScI), which is operated by the Association of Universities for Research in Astronomy (AURA), Inc., under NASA contract NAS 5-03127. A.Z. acknowledges support by the Israel Science Foundation Grant No. 864/23, E.Z. acknowledges project grant 2022-03804 from the Swedish Research Council (Vetenskapsr\aa{}det). H.A. acknowledge support from CNES, focused on the JWST mission, and the French National Research Agency (ANR) under grant ANR-21-CE31-0838.

The raw JWST data used in this work are publicly available on the \texttt{Barbara A. Mikulski Archive for Space Telescopes} (\texttt{MAST}). These observations specifically are associated with the JWST GO program numbers 3293, and 1840, as well as JWST DDT program number 9223. The HST data used here are also available on the \texttt{MAST}, under program numbers 14037, 15117, and 15936. All data used in this work can be found on the \texttt{MAST} under the following DOI: \url{10.17909/nwc2-d542}.
\end{acknowledgments}

%\begin{contribution}
%[...].
%\end{contribution}

\facilities{This work is based on observations obtained with the NASA/ESA/CSA JWST and the NASA/ESA HST, retrieved from the \texttt{MAST} at the STScI.}

\software{This research made use of \texttt{Astropy}, a community-developed core Python package for Astronomy \citep[][]{astropy13,astropy18} as well as the packages \texttt{NumPy} \citep{vanderwalt11}, \texttt{SciPy} \citep{virtanen20}, \texttt{Matplotlib} \citep{hunter07} and the \texttt{MAAT} Astronomy and Astrophysics tools for \texttt{MATLAB} \citep{maat14}. We further used the \texttt{photutils} \citet{photutils_v2.2.0}, \texttt{specutils} \citep[][]{specutils21}, and \texttt{emcee} \citep{foreman-mackey13} packages.}


\appendix

\section{Photometric measurements} \label{app:photomety-table}

%\begin{rotatetable}
\begin{deluxetable*} {lccccccccccccc}
\tablecaption{Customized aperture photometry of our lensed star candidates as described in appendix~\ref{app:photomety-table}. \label{tab:photometry}}
\tablehead{
\colhead{ID} & \colhead{F435W} & \colhead{F200LP} & \colhead{F350LP} & \colhead{F814W} & \colhead{F090W} & \colhead{F115W} & \colhead{F150W} & \colhead{F200W} & \colhead{F277W} & \colhead{F356W} & \colhead{F410M} & \colhead{F444W} & \colhead{F480M}}
\startdata
70878   &   $<2$    &   $<1$    &   $<1$    &   $<1$    &   $<0.5$    & $<0.4$  &   $<0.6$      &   $<0.5$      &   $2.9\pm0.4$ &   $7.3\pm0.3$ &   $9.2\pm0.8$ &   $10.6\pm0.5$    &   $9.1\pm1.3$\\
43258   &   $<2$    &   $<1$    &   $<1$    &   $<2$    &   $<0.6$    & $<0.8$  &   $0.9\pm0.6$ &   $5.1\pm0.9$ &   $4.5\pm0.7$ &   $5.6\pm0.3$ &   $6.8\pm1.0$ &   $4.9\pm0.4$     &   $4.2\pm1.5$\\
24951   &   $<2$    &   $<1$    &   $<1$    &   $<1$    &   $<0.4$    & $<0.4$  &   $0.9\pm0.5$ &   $1.4\pm0.5$ &   $3.5\pm0.4$ &   $5.4\pm0.3$ &   $5.5\pm0.5$ &   $5.9\pm0.3$     &   $3.2\pm1.3$\\
68501   &   $<2$    &   $<1$    &   $<1$    &   $<2$    &   $<0.5$    & $<0.5$  &   $<0.7$      &   $<0.7$      &   $1.2\pm0.4$ &   $3.4\pm0.4$ &   $3.9\pm0.5$ &   $5.0\pm0.4$     &   $2.3\pm1.1$\\
\enddata
%\tablenotetext{a}{}
\tablecomments{Fluxes are given in nJy. Quoted upper limites represent $1\sigma$ limits.}
\end{deluxetable*}
%\end{rotatetable}

We use the \texttt{photutils} package \citep[\texttt{v2.2.0};][]{photutils_v2.2.0} to measure photometry in all available and deep enough ($>30$\,magnitudes) filters, i.e.\ JWST/NIRCam F090W, F115W, F150W, F200W, F277W, F356W, F410M, F444W, and F480M, HST/ACS F435W, and F814W, and HST/WFC3-UVIS F200LP, and F350LP (see section~\ref{sec:data}). The photometry is measured in circular apertures of 0.2\arcsec\ diameter from the BCG-subtracted GLIMPSE mosaics after subtracting a 20\,pixel median-filtered background image from each mosaic. We apply an aperture correction derived from the growth curve of the point-spread-function (PSF) in each band. The photometric uncertainties are measured by placing empty apertures of the same size in the vicinity of each source, using the GLIMPSE catalog segmentation map to make sure that there is no object present, and computing the standard deviation of the measured fluxes. In the case of ID~43258 (\textit{Hedorah}), we note significant contamination by the host arc (see Fig~\ref{fig:observations}). In order to correct for that, we measure fluxes in same size apertures either side of the object and subtract the median of these two from the object flux. The resulting fluxes are listed in Tab.~\ref{tab:photometry} and we show the SED of each object in Fig.~\ref{fig:observations}.

The surface brightness of the host-arcs adjacent to the stars is measured in a similar manner to the background estimate for \textit{Hedorah}: We place apertures on the expected arc either side of the lensed stars, and use the average fluxes of each arc to infer a surface brightness by dividing by the are of the aperture. This is done without applying the median filter that we used for the point-source photometry in order to not over-subtract any diffuse extended arc emission. The host arcs of the three AGB star candidates (IDs~70878, 24951, and~68501) are not detected ($<2\sigma$), yielding $5\sigma$ upper limits of $\mathrm{SB}<21\,\mathrm{nJy}\,\mathrm{arcsec}^{-2}$ in F115W, and $\mathrm{SB}<25\,\mathrm{nJy}\,\mathrm{arcsec}^{-2}$ in F444W assuming GLIMPSE depths \citep[][]{atek25}. The host arc of ID~43258, \textit{Hedorah}, is detected, as can also be seen in Figs.~\ref{fig:observations} and~\ref{fig:hedorah-host_spec}, with $\mathrm{SB}=29\pm8\,\mathrm{nJy}\,\mathrm{arcsec}^{-2}$ in both filters.

To test whether the lensed star candidates could be transients, we apply the same photometry measurement method as above to the JWST/NIRCam imaging from GO-1840, which were obtained one year prior to GLIMPSE, as well as the HST/WFC3-IR data from the HFF (see section~\ref{sec:data}). In conjunction with visual inspection, this yields two marginal detections, of ID~70878 and ID~68501, both in the F444W-band, at signal-to-noise ratios of $2.05$ and $5.28$, respectively. The measured F444W-band magnitudes are $m_{\mathrm{F444W}}=29.34\pm0.53$ and $m_{\mathrm{F444W}}=28.27\pm0.21$. All four lensed star candidates remain un-detected in the other GO-1840 NIRCam bands, as well as all WFC3-IR bands. This is however not surprising since these data are much shallower than GLIMPSE and do not achieve sufficient depth to robustly detect the fluxes listed in Tab.~\ref{tab:photometry}. We therefore conclude that at least two of the lensed star candidates presented in this work could be persistent, but cannot rule-out a transient nature with the present data.

\section{Spectroscopic redshift measurement of \textit{Hedorah}'s host arc} \label{app:spectroscopy}
The GLIMPSE-DDT NIRSpec program (program ID: 9223; see section~\ref{sec:data}) includes image~401.2 of the host-arc of ID~43258, the yellow super-giant `\textit{Hedorah}', which enables us to for the first time measure its spectroscopic redshift. As can be seen in Fig.~\ref{fig:hedorah-host_spec}, the MSA slitlet includes two sources, image~401.2 from \textit{Hedorah}'s host arc, and image~13.2 from a known multiple-image system in AS1063 that was detected with HST and confirmed Multi Unit Spectroscopic Explorer \citep[MUSE;][]{bacon10} on ESO's Very Large Telescope (VLT) at $z_{\mathrm{spec}}=4.11$ \citep[e.g.][]{bergamini19,richard21}. We disentangle the two spectra by using a custom nodded background subtraction in the spectroscopic reduction. In this custom subtraction, when combining nodded background frames we exclude frames where the spectral trace of the neighboring source would contaminate the primary source. While this reduces the overall depth of the combined background frames, the benefits from the lack of contamination outweigh this minor depth reduction, especially for a strong line-emitting object. In this case, the known redshift of image~13.2 enables us to easily select the contaminating source and isolate the spectrum of image~401.2.

This cleaned spectrum of image 401.2 shows a blue continuum as well as one strong and prominent emission line (see Fig.~\ref{fig:hedorah-host_spec}). While it is typically difficult to infer spectroscopic redshift from one emission line alone, we in this case in particular can take advantage of the SL geometry: With numerous spectroscopically confirmed multiple-image systems in the vicinity, our SL model enables us to place a geometric prior on system 401, $z_{\mathrm{geo}}=4.0_{-0.2}^{+0.3}$. This makes the strong emission line most likely H$\alpha$. We fit the line and continuum using \texttt{specutils} \citep[][]{specutils21} and measure a spectroscopic redshift of $z_{\mathrm{spec}}=3.7152\pm0.0001$ for system 401 and its lensed star \textit{Hedorah}, as shown in Fig.~\ref{fig:hedorah-host_spec}.

\section{Strong lensing model} \label{app:SL}
We here construct a new parametric SL model of AS1063 for GLIMPSE, using the python version the \citet{zitrin15a} analytic method, \texttt{AstroLensPy} \citep[][]{allingham26}. The model comprises two smooth cluster-scale DM halos parametrized as pseudo-isothermal elliptical mass distributions \citep[PIEMDs;][]{kassiola93} and 303 cluster member galaxies parametrized as dual pseudo-isothermal ellipsoids \citep[dPIEs;][]{eliasdottir07}. The second cluster PIEMD halo is centered on a group of galaxies in the north-east of the main cluster and has been found necessary to correctly model the mass distribution of AS1063 \citep[][]{bergamini19,beauchesne24}. We use a total of 91 multiple images belonging 34 sources as constraints. Out of these, 31 systems have spectroscopic redshifts measured with MUSE \citep[][]{bergamini19,richard21,beauchesne24}. We further measure 5 new spectroscopic redshifts of multiple image systems from the JWST/NIRSpec DDT-9223 (see section~\ref{sec:data}), including the host-arc of \textit{Hedorah} (see appendix~\ref{app:spectroscopy} and Fig.~\ref{fig:hedorah-host_spec}), and for the first time including 2 multiple-image systems in the north-eastern substructure which was previously unconstrained. The SL model is optimized in the source plane with 8000 steps of a Monte-Carlo Markov Chain (MCMC) analysis run with \texttt{emcee} and achieves a final average lens-plane image reproduction error of $\Delta_{\mathrm{RMS}}=0.32\arcsec$. For more information on the SL model, we defer the reader to the upcoming dedicated paper, Furtak at al. (in prep.).

%% For this sample we use BibTeX plus aasjournalv7.bst to generate the

\bibliography{references}
\bibliographystyle{aasjournalv7}


\end{document}